**The Renaissance of Main Belt Asteroid Science**

Simone Marchi
Southwest Research Institute, 1050 Walnut St, Boulder, CO 80302, USA
marchi@boulder.swri.edu

Carol A. Raymond
Jet Propulsion Laboratory, California Institute of Technology, Pasadena, CA, USA
carol.a.raymond@jpl.nasa.gov

Christopher T. Russell
University of California, Los Angeles, 603 Charles Young Drive, Los Angeles, CA 90095, USA
ctrussel@igpp.ucla.edu

1. **Introduction.**

More than 200 years separate us from the scientific turmoil generated by the discovery of Ceres, the largest Main Belt asteroid, in a region of the Solar System that was expected to harbor a planet. Yet scientists are still wrestling with fundamental questions such as: How did asteroids form? What are they made of? What do they tell us about the evolution of our Solar System? These questions are at the core of modern planetary science, and are drivers for the robotic exploration of the Solar System.

Asteroid science progressed at a slow pace in the decades after Ceres' first observations. As more asteroids were discovered in the 19$^{th}$ and 20$^{th}$ century, their formation and physical properties remained largely unknown, but one thing became immediately clear: asteroids are tiny compared to planets. So, one may ask, are asteroids just a minor detail in the grand scheme of the Solar System?

In the 1970s, technology finally caught up and sophisticated telescopic observations became possible (e.g., McCord et al. 1970). Around the same time, it was established that asteroids are the likely source of meteorites found on Earth (Arnold 1965; Wetherill 1967), allowing a first quantitative understanding of their nature based on meteorite properties. The first serendipitous flyby of Main Belt asteroid Gaspra in 1991 by the NASA *Galileo* spacecraft enroute to Jupiter was a pivotal moment in planetary sciences (Belton et al. 1992), at which asteroids became resolved astronomical objects like the Moon and Mars. As more data from telescopic observations, theoretical models, and space exploration started to accumulate, it became clear that the Main Belt was a fundamental part of the structure of our Solar System, a natural divide between the inner rocky terrestrial planets and the outer gaseous planets.

This book depicts a vivid and vibrant image of modern Main Belt asteroid science. In the last decade, thanks to the exploration by the NASA *Dawn* mission and advent of high-resolution Earth-bound observations (Vernazza et al. 2021; Russell and Rayman 2021; see chapters 1, 2), we have entered a Renaissance of Main Belt asteroid science. Formation theories, dynamical models, meteorite geochemical data, remote and *in-situ* observations synergistically show asteroids are leftover building blocks of planetary formation and tracers of important evolutionary processes (e.g., collisions, orbital migration) that have shaped the evolution of the early Solar System. And, perhaps, asteroids will be exploited for their resources in the future.

## 2. Main Belt asteroid science state-of-the-art.

Looking back at the progress of Main Belt asteroid science reveals the scientists' constant struggle to make sense of new information. An early theory proposed by Wilhelm Olbers in 1802 envisioned that asteroids formed by the explosion of a larger planet (e.g., see Cunningham and Orchiston 2013). This view quickly became obsolete when more asteroids were discovered to span a wide range of heliocentric distances and orbital inclinations. In addition, the cumulative mass of the Main Belt is very low compared to nearby Mars and Jupiter implying a discontinuity in the distribution of material in the proto-solar nebula (Kuiper 1956). The broad orbital distribution of asteroidal eccentricities and inclinations were explained via self-stirring by sizable planetary embryos embedded in the Main Belt, which would have subsequently been removed together with 99.9% of the original Main Belt mass (Wetherill 1992). Or, perhaps, the Main Belt was never so massive, and the asteroid's orbital stirring was caused by external processes, such as dynamical perturbations from nearby Mars and Jupiter (Ward 1981).

To make the picture even fuzzier, spectroscopic observations of asteroids and meteoritical analyses indicate a large compositional variability in the Main Belt (Vernazza et al. 2021; see chapter 1). Yet, isotopic data from meteorites favor the existence of two non-contiguous regions of asteroid formation, arguably the terrestrial planet region and the Jupiter-Saturn region (Bermingham and Kruijer 2021; see chapter 14). Recent dynamical models predict that the Main Belt could have started off (nearly) empty, and was populated at a later time due to planetesimal-planet orbital scattering (Raymond and Nesvorny 2021; see chapter 15). In this view, the Main Belt is a receptor of planetesimals from a wide range of heliocentric distances.

To examine this idea more closely, Figure 1 shows the distribution of Main Belt asteroids larger than 100 km (diameter), and their spectral types. These objects are considered to be primordial, that is, planetesimals that are not fragments of larger parent bodies and therefore better represent their distribution in the Main Belt, while smaller asteroids are thought to be fragments of collisional evolution of larger parent bodies (Bottke et al 2005; Morbidelli et al 2009). The spectral types of asteroids are proxy for their compositions, as inferred by comparison with meteorites (Vernazza et al. 2021; see chapter 1). The figure shows that pristine, volatile-rich asteroids represented by the B, C, and D types are preferentially found in the outer Main Belt, leading to the idea of a gradual increase of volatiles (and water) with heliocentric distance. At a closer look, however, several notable outliers are evident: a few large pristine asteroids (e.g., Nemausa, Fortuna, Zelinda) are found in the inner Main Belt amidst volatile-poor asteroids (e.g., Vesta), and vice versa in the outer Main Belt. Scattered throughout the Main Belt, elusive X-type asteroids have been linked to both highly evolved bodies (e.g., rich in metal such as Psyche) to primitive objects (similar to CM carbonaceous meteorites; Fornasier et al. 2011). It is difficult to imagine that asteroids with such wildly different compositions may have formed close to each other as reflected in their current orbital arrangement. These observations seem to indicate, instead, that the Main Belt is the result of widespread mixing. *Dawn's* detailed observations at Vesta and Ceres have contributed to further refinements of these arguments, and at the same time have unveiled unexpected details.

Vesta is volatile-poor and the howardite, eucrite and diogenite (HED) meteorites that show evidence of being sourced from Vesta constrain its bulk water content to have been a few percent at most (McSween and Binzel 2021; Toplis and Breuer 2021; see chapters 3, 4). On the contrary, Ceres has a bulk water content up to 30-40% (Castillo-Rogez and Bland 2021;

Ermakov and Raymond 2021; see chapters 11, 12). Thus, the two most massive objects in the Main Belt comprising about 45% of the total mass have water content that differs by an order of magnitude, yet their semi-major axes are separated by only about 0.4 AU (indeed, Vesta and Ceres orbits intersect each other). Such organization would require the "fine tuning" of the location of the water condensation line in the protoplanetary disk to fall between Vesta (~2.36 AU) and Ceres (~2.76 AU). This fine tuning, however, is contrary to our understanding of how the protoplanetary disk evolved. For instance, the disk radial temperature profile is expected to dramatically migrate radially during the first Myrs of evolution due to viscous dissipation (e.g., Bitsch et al. 2015). Therefore, it is unclear how disk evolution could have resulted in a sharp compositional difference between Vesta and Ceres. In addition, ammonia may have been an important raw ingredient of the materials from which Ceres formed (Ammannito and Elhmann 2021; see chapter 9). If one assumes that the condensation of super volatile ammonia hydrate (~80 K) occurred at Ceres' present heliocentric distance, then the entire Main Belt would have been within the water condensation zone (~180 K), contrary to the presence of many volatile-poor asteroids.

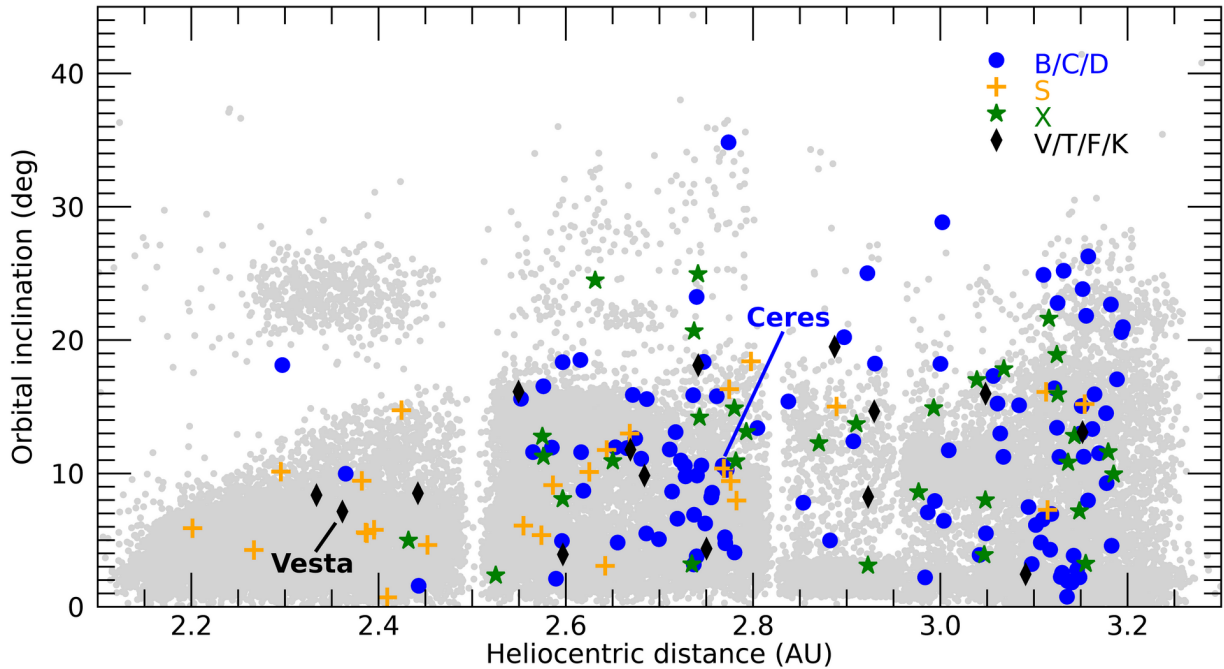

*Figure 1.* *Distribution of inclination vs heliocentric distance (proper semi-major axis) for Main Belt asteroids (small light gray dots). Asteroids larger than 100 km in diameter are indicated by their spectral types: B/C/D represents volatile-rich objects, S indicates volatile-poor objects, X are not well characterized, and could range from metal-rich to more pristine assemblages; V/T/F/K are peculiar compositions typically interpreted as highly evolved (e.g., Vesta is a V-type).*

The *Dawn* mission showed beyond any reasonable doubt that most (if not all) of the HED meteorites originate from Vesta based on its surface composition (Combe and Yamashita 2021;

see chapter 6), and geomorphology (Buczkowski et al. 2021; see chapter 5), including the presence of two large impact basins near the south pole (Bottke and Jutzi 2021; see chapter 16). The Vesta-HED connection is of particular importance for our understanding of the early Solar System evolution, as it allowed a full closure of modeling, remote observations and sample analysis, which has been previously possible only for the Moon and Mars (e.g., Marchi et al. 2013).

Collectively, these arguments suggest that the majority of large Main Belt asteroids, including Vesta and Ceres, formed elsewhere in the Solar System. If so, where did they form?

Dynamical models (Raymond and Nesvorny 2021; see chapter 15) show that there are multiple pathways by which planetesimals from the terrestrial planet regions (1-2 AU) and the outer solar system (>5 AU) could end up in the Main Belt. This double implantation could also explain isotopic evidence for two distinct reservoirs (inner vs outer Solar System), potentially due to a barrier resulting from the formation of Jupiter that would have temporarily limited exchange of materials between the two reservoirs (Bermingham and Kruijer 2021; see chapter 14). While this view of a divide in the formation zones of planetesimals and their subsequent radial mixing is rapidly gaining consensus, open issues remain. Consider the following examples.

Elemental and isotopic data from HEDs seem to require that Vesta accreted from material that was ¾ H chondrites and ¼ CM chondrites, within 1-2 Myr after solar system formation (Toplis and Breuer 2021; see chapter 4). Interestingly, H and CM chondrites belong to the two distinct meteorite reservoirs based on isotopic data, implying that mixing of planetesimals between the two reservoirs occurred very early in the Solar System. However, CM meteorites are thought to have accreted between 2.5 and 3.5 Myr (Kruijer et al. 2020).

Ceres could have accreted beyond Jupiter to account for its ammonia and high carbon content (Ammannito and Ehlmann 2021; Prettyman et al. 2021; see chapters 8, 9). It is therefore possible that Ceres accreted in the same reservoir as other carbonaceous chondrites, but if so, it is unclear why these meteorites have lower concentrations of ammonia and carbon than Ceres. Another interesting aspect is how Ceres grew to its current size. Formation models envision planetesimals formed due to the collapse of pebble clouds with a final preferred size of ~100 km (Klahr et al. 2021; see chapter 13). In this scenario, subsequent growth via planetesimal collisions or pebble accretion is required. It could be that this second stage of accretion is responsible for Ceres's compositional differences from carbonaceous chondrites. Ceres could be one of many 1000-km objects that formed in this way and ended up in the Main Belt, while its siblings remain in the outer Solar System.

There is also another fundamental consequence of the different bulk volatile content between Vesta and Ceres. The presence of water, in particular, has the capability to limit the rise of internal temperature powered primarily by $^{26}$Al, $^{60}$Fe radioactive decay. As result, Vesta reached internal temperatures that were high enough for metal-silicate differentiation (Toplis and Breuer 2021; see chapter 4), while Ceres' internal temperatures remained low enough to stunt large-scale metal-silicate differentiation (Castillo-Rogez and Bland 2021; see chapter 11). As a result of this radically different evolution these objects evolved on divergent paths, as inferred by their internal structure (Ermakov and Raymond 2021; see chapter 12), surface composition (Combe and Yamashita 2021; De Sanctis and Raponi 2021; see chapters 6, 7), and geomorphology (Buczkowski et al 2021; Williams et al. 2021; see chapters 5, 10). Still, there remain many uncertainties to precisely understand the chemical pathways that led to Ceres' complex surface composition, and its possible recent geological activity (Raymond et al. 2020).

### 3. Looking ahead.

The 21st-century Renaissance of Main Belt asteroid science reflects the fact that asteroids hold crucial and unique clues about the formation and evolution of our Solar System. But as in an unfinished painting of Renaissance master Leonardo da Vinci, there is more left for us to imagine, explore and eventually discover. Planned missions such as NASA's *Lucy* and *Psyche* (scheduled to launch in 2021 and 2022) will surely provide additional colorful strokes to our ever-evolving portrait of the Main Belt. *Psyche* will expand Main Belt asteroid exploration to the uncharted territory of metal-rich objects, while *Lucy* will study unexplored Trojan asteroids, a class of objects orbiting the Sun in Jupiter's stable Lagrangian points. Both missions will provide new detailed information about the formation of the Solar System, including asteroid migration and implantation processes that could have been important for the Main Belt.

The *Psyche* and *Lucy* missions will hopefully pave the way for a new generation of bold Main Belt-oriented missions. These missions should not be limited to remotely observing asteroids (flybys or rendezvous), but should also sample their surfaces for in-situ chemical and mineralogical analysis, or bring those samples back to Earth. It is generally considered that the ensemble of known meteorites are samples of roughly 80-100 distinct parent bodies (Greenwood et al. 2020), a number that is 50% of primordial asteroids in the Main Belt. This suggests that our understanding of the building blocks of the Solar System is limited and potentially strongly biased by the few meteorites in our labs. Synergy between spacecraft exploration of large planetesimals, analysis of returned samples, and advances in measuring isotopic characteristics of individual components of meteorite samples promises to accelerate our understanding of the many clues held within the Main Belt asteroids that still await discovery.

As this book goes to press, the JAXA *Hayabusa-2* mission has returned to Earth samples from the small near-Earth asteroid Ryugu, and NASA's *OSIRIS-Rex* is ahead back to Earth with its precious cargo of samples from Bennu. These small asteroids originate from the Main Belt, as likely fragments of larger parent bodies. We see these ambitious missions as precursors of similar efforts to land on and sample large, Main Belt asteroids. It is the authors' belief that only with these future investigations will the portrait of the Main Belt be, hopefully, completed.

### References.